# Power-Balanced Orthogonal Space-Time Block Code

Chau Yuen, Yong Liang Guan and Tjeng Thiang Tjhung, *Member, IEEE*

*Abstract*—In this paper, we propose two new systematic ways to construct amicable orthogonal designs (AOD), with an aim to facilitate the construction of power-balanced orthogonal space-time block codes (O-STBC) with favorable practical attributes. We also show that an AOD can be constructed from an Amicable Family (AF), and such a construction is crucial for achieving a power-balanced O-STBC. In addition, we develop design guidelines on how to select the "type" parameter of an AOD so that the resultant O-STBC will have better power-distribution and code-coefficient attributes. Among the new O-STBCs obtained, one is shown to be optimal in terms of power distribution attributes. In addition, one of the proposed construction methods is shown to generalize some other construction methods proposed in the literature.

## I. INTRODUCTION

AN Orthogonal Space-Time Block Code (O-STBC) can provide full transmit diversity with simple linear decoding. Due to these advantages, O-STBC has drawn a lot of research attention. In [1-3], the Alamouti STBC is generalized to square O-STBC for more than two transmit antennas by using different mathematical techniques, while in [4,5], non-square O-STBC are investigated in an attempt to increase the maximum achievable code rate of O-STBC. Nonetheless, square O-STBC has the advantages of minimal decoding delay and applicability for differential modulation.

Some of the square O-STBCs, such as the ones in [2,3], contain "zero" symbols; while those in [1] have irrational numbers in the code coefficients. The regular transmission of "zeros" implies turning off the transmit antennas at regular intervals. This leads to an undesirable low-frequency interference and some difficulty in the front-end power amplifier design [6-8]. Therefore, the problem of designing power-balanced O-STBC with less or no zero symbols has been investigated in [6,8-11].

Manuscript received September 25, 2007.
C. Yuen is with the Institute for Infocomm Research, Singapore (phone: 65-68745679; fax: 65-67744990; e-mail: cyuen@i2r.a-star.edu.sg).
Y. L. Guan is with the Department of Electrical and Electronic Engineering, Nanyang Technological University, Singapore (e-mail: eylguan@ntu.edu.sg).
T. T. Tjhung is with the Institute for Infocomm Research, Singapore (e-mail: tjhungtt@i2r.a-star.edu.sg).

In [6], a method is proposed such that the transmission power of existing O-STBC is distributed as equally as possible between different antennas; while in [8-11], other algebraic techniques such as Williamson and Wallis-Whiteman matrices are used to construct power-balanced O-STBC. Besides the zero symbols problem, irrational numbers in the code coefficients require floating-point multiplications in both the transmitter and receiver of the O-STBC system. This is inconvenient to implement compared to having just 1 in the code-coefficients, which requires only simple additions/subtractions.

In this paper, we focus on the construction of O-STBC's from amicable orthogonal designs (AOD) that avoids zero and irrational coefficients. We propose two new systematic methods to construct higher-order AOD's or Amicable Family's (AF) from lower-order AOD's or AF's, and use them to construct new square O-STBCs. We then investigate the relationship between the "type" parameter of the AOD's and the power-distribution and code-coefficient characteristics of the constructed O-STBC's. We will evaluate the advantages of the newly constructed O-STBCs over the existing ones. We will also give a comparison of the proposed construction methods to the methods in [9], and show that the proposed construction methods can be generalized to the latter.

## II. ORTHOGONAL STBC

A linear STBC, G, can be represented as:

$$\mathbf{G} = \sum_{i=1}^{k}(x_i^R \mathbf{A}_i + jx_i^I \mathbf{B}_i) \quad (1)$$

where $\mathbf{A}_i$ and $\mathbf{B}_i$ are the "dispersion matrices" (both of dimension $p \times n_t$), where $x_i^R$ and $x_i^I$ represent respectively, the real and imaginary parts of the $i^{th}$ transmitted symbol, $p$ is the code length, $n_t$ is the number of transmit antennas, and $k$ is the number of complex symbols being transmitted over $p$ periods of time. Hence the code rate of a STBC is $k/p$.

For an O-STBC, its dispersion matrices, $\mathbf{A}_i$ and $\mathbf{B}_i$, must satisfy the following constraints [2]:

(i) $\mathbf{A}_i^H \mathbf{A}_i = \mathbf{I}_{n_t}$ , $\mathbf{B}_i^H \mathbf{B}_i = \mathbf{I}_{n_t}$   $1 \le i \le k$

(ii) $\mathbf{A}_i^H \mathbf{A}_q = -\mathbf{A}_q^H \mathbf{A}_i$ , $\mathbf{B}_i^H \mathbf{B}_q = -\mathbf{B}_q^H \mathbf{B}_i$  $1 \le i \ne q \le k$   (2)

(iii) $\mathbf{A}_i^H \mathbf{B}_q = \mathbf{B}_q^H \mathbf{A}_i$   $1 \le i, q \le k$

where the superscript H represents Hermitian (i.e. conjugate transpose) of a matrix. In order to achieve full diversity, $p$ has to be greater or equal to $n_t$. Hence a square STBC design with $p = n_t$ gives the minimum possible code length. In addition, a square design can be applied to differential unitary space-time coding. So we only consider the design of square O-STBC in this paper.

To construct a square O-STBC, one can make use of the AOD. Let us first review the relationship between AOD's and O-STBC's [2]:

*Definition 1* [12]: Let the matrices $\mathcal{A} = \mathbf{A}_1 a_1 + \ldots + \mathbf{A}_s a_s$ and $\mathcal{B} = \mathbf{B}_1 b_1 + \ldots + \mathbf{B}_t b_t$ be orthogonal designs of the same *order* $n$ (i.e. both $\mathcal{A}$ and $\mathcal{B}$ are $n \times n$), where $\mathcal{A}$ is of "*type*" $(f_1, \ldots, f_s)$ on the variables $\{a_1, \ldots, a_s\}$ and $\mathcal{B}$ is of "*type*" $(g_1, \ldots, g_t)$ on the variables $\{b_1, \ldots, b_t\}$. $\mathcal{A}$ and $\mathcal{B}$ are said to be *Amicable Orthogonal Design* (AOD) if

$$\mathcal{A}^H \mathcal{B} = \mathcal{B}^H \mathcal{A} \quad (3)$$

A necessary and sufficient condition for an AOD of "*type*" $(f_1, \ldots, f_s; g_1, \ldots, g_t)$, as defined in *Definition 1*, to exist is that there exists a family of matrices $\{\mathbf{A}_1, \ldots, \mathbf{A}_s; \mathbf{B}_1, \ldots, \mathbf{B}_t\}$ satisfying:

(0) $\mathbf{A}_i * \mathbf{A}_l = \mathbf{0}$    $1 \le i \ne l \le s$ ;
$\mathbf{B}_q * \mathbf{B}_m = \mathbf{0}$    $1 \le q \ne m \le t$

(i) $\mathbf{A}_i^H \mathbf{A}_i = f_i \mathbf{I}_n$    $1 \le i \le s$ ;
$\mathbf{B}_q^H \mathbf{B}_q = g_q \mathbf{I}_n$    $1 \le q \le t$    (4)

(ii) $\mathbf{A}_i^H \mathbf{A}_l + \mathbf{A}_l^H \mathbf{A}_i = \mathbf{0}$    $1 \le i \ne l \le s$ ;
$\mathbf{B}_q^H \mathbf{B}_m + \mathbf{B}_m^H \mathbf{B}_q = \mathbf{0}$    $1 \le q \ne m \le t$

(iii) $\mathbf{A}_i^H \mathbf{B}_q - \mathbf{B}_q^H \mathbf{A}_i = \mathbf{0}$    $1 \le i \le s, 1 \le q \le t$

where $\mathbf{A}_i$ and $\mathbf{B}_q$ consist of only $\{0, 1, -1\}$, the symbol $*$ represents the Hadamard product, and $\mathbf{0}$ is a zero matrix. ∎

By comparing the constraints in (4) for the AOD matrix with that in (2) for the O-STBC dispersion matrix, it can be seen that the $\mathbf{A}_i$ and $\mathbf{B}_q$ matrices defined in (4) can be used as the dispersion matrices of an O-STBC since they satisfy the design constraints in (2). The "*order*" parameter $n$ of an AOD corresponds to the number of transmit antennas $n_t$ of the O-STBC, while the code length $p$ of the constructed O-STBC will be equal to $n$, since an AOD is a square design. In addition, the "*type*" parameter of an AOD, i.e. $(f_1, \ldots, f_s; g_1, \ldots, g_t)$, is related to the power distribution of the transmit symbols of an O-STBC. This will be elaborated later in Section IV. Furthermore, the total number of variables, i.e. $s + t$, of the AOD represents the number of real symbols, i.e. $2k$, carried by the O-STBC. Hence the code rate of an O-STBC constructed from an AOD will be $(s + t) / 2n$. It has been shown in [1-3] that the maximal code rate of a square O-STBC is ¾ for four transmit antennas and ½ for eight transmit antennas by using the following property of an AOD:

*Lemma 1* [12]: For an AOD of order $n$, where $n = 2^a b$, $a$ and $b$ are both integers and $b$ is odd, the total number of variables in an AOD (i.e. $s + t$) is upper bounded by $2a + 2$, and that bound is achieved. ∎

*Definition 2* [12]: An *amicable family* (AF) of type $(f_1, \ldots, f_s; g_1, \ldots, g_t)$, of order $n$ is a collection of matrices $\{\mathbf{A}_1, \ldots, \mathbf{A}_s; \mathbf{B}_1, \ldots, \mathbf{B}_t\}$ satisfying (4)(i), (ii), and (iii), but not the "disjointness statement" (4)(0). ∎

It will be shown that an AF plays an important role in the construction of power-balanced O-STBC's.

### III. CONSTRUCTION OF AOD

We now propose two new methods to construct higher-order AOD's from lower-order AOD's. In the first method we construct an AOD of order $4n$ from an AOD of order $n$, while in the second method we construct an AOD of order $2n$ from an AOD of order $n$. We shall show that the resultant higher-order AOD achieves the maximum number of variables (and its associated O-STBC achieves the maximum code rate) if the lower-order AOD used to generate it achieves the maximum number of variables.

*Construction 1*: If $\{\mathbf{A}_i, 1 \le i \le s; \mathbf{B}_q, 1 \le q \le t\}$ is an AOD / AF of order $n$ with $s + t$ variables of "*type*" $(f_1, \ldots, f_s; g_1, \ldots, g_t)$, then

$$\{\mathbf{B}_1 \otimes \mathbf{M}_1, \mathbf{B}_1 \otimes \mathbf{M}_2, \mathbf{B}_1 \otimes \mathbf{M}_3, \mathbf{A}_i \otimes \mathbf{I}_4, 2 \le i \le s;$$
$$\mathbf{A}_1 \otimes \mathbf{N}_1, \mathbf{A}_1 \otimes \mathbf{N}_2, \mathbf{A}_1 \otimes \mathbf{N}_3, \mathbf{B}_q \otimes \mathbf{I}_4, 2 \le q \le t\}$$

is an AOD / AF of order $4n$ with $s + t + 4$ variables and "*type*" $(g_1, g_1, g_1, f_2, \ldots, f_s; f_1, f_1, f_1, g_2, \ldots, g_t)$, where $\otimes$ is the Kronecker product, if the matrices $\mathbf{M}_i$ and $\mathbf{N}_i$, for $1 \le i \le 3$, satisfy the following conditions:

(0) $\mathbf{M}_i * \mathbf{M}_q = \mathbf{0}$,   $\mathbf{N}_i * \mathbf{N}_q = \mathbf{0}$    $1 \le i \ne q \le 3$

(i) $\mathbf{M}_i^H \mathbf{M}_i = u_i \mathbf{I}_4$,   $\mathbf{N}_i^H \mathbf{N}_i = v_i \mathbf{I}_4$    $1 \le i \le 3$

(ii) $\mathbf{M}_i^H \mathbf{M}_q + \mathbf{M}_q^H \mathbf{M}_i = \mathbf{0}$,   $\mathbf{N}_i^H \mathbf{N}_q + \mathbf{N}_q^H \mathbf{N}_i = \mathbf{0}$   $1 \le i \ne q \le 3$    (5)

(iii) $\mathbf{M}_i^H \mathbf{N}_q - \mathbf{N}_q^H \mathbf{M}_i = \mathbf{0}$,    $1 \le i, q \le 3$

(iv) $\mathbf{M}_i * \mathbf{I}_4 = \mathbf{0}$,   $\mathbf{N}_i * \mathbf{I}_4 = \mathbf{0}$,    $1 \le i \le 3$

(v) $\mathbf{M}_i^H + \mathbf{M}_i = \mathbf{0}$,   $\mathbf{N}_i^H + \mathbf{N}_i = \mathbf{0}$,   $1 \le i \le 3$

Proof of *Construction 1*: The proofs are routine, hence omitted due to space constraint. ∎

The following observations can be made from *Construction 1*:
- An interesting observation is that since the conditions (5)(0-iii) are equivalent to the conditions (4)(0-iii), it implies that $\{\mathbf{M}_1, \mathbf{M}_2, \mathbf{M}_3; \mathbf{N}_1, \mathbf{N}_2, \mathbf{N}_3\}$ must themselves be an AOD of order 4 with six variables and type $(u_1, u_2, u_3; v_1, v_2, v_3)$.
- If the target design is an AF, then conditions (5)(0) and (5)(iv) in *Construction 1* can be neglected because they are related to the "disjointness statement". On the other hand, if the target design is an AOD, then $u_i$ and $v_i$ must both be 1 for $1 \le i \le 3$. This is because in an AOD of order 4, if $u_i$ or $v_i$ is greater than 1, the "disjointness statement" (5)(0) and (5)(iv) will be violated.
- If the lower-order design is an AOD, the resultant higher-order design will also be an AOD. However if the lower-order design is an AF, the resultant higher-order design may be an AOD or an AF. For an illustration, an AOD of order 8 constructed from an AF of order 2 will be shown later in *Example 1*.

*Proposition 1*: An AOD of order $4n$ constructed by *Construction 1* can achieve the maximum number of variables if an AOD of order $n$ with the maximum number of variables is used to construct it.

Proof of *Proposition 1*: From *Lemma 1*, the maximum number of variables for an AOD of order $n$ is bounded by $2a + 2$, where $n = 2^a$. This implies that the maximum number of variables for an AOD of order $4n$ is bounded by $2a + 6$. Therefore the difference between the maximum number of variables of an AOD of order $n$ and an AOD of order $4n$ is at most 4. From *Construction 1*, the number of variables of an AOD of order $4n$ is $s + t + 4$. This is also 4 more than the number of variables of an AOD of order $n$ used to generate it. Hence if an AOD of order $n$ that achieves the maximum number of variables is used, an AOD of order $4n$ that achieves the maximum number of variables can be constructed. ∎

Examples of $\{\mathbf{M}, \mathbf{N}\}$ that satisfy the conditions in (5) are:

$$\mathbf{M}_1 = \begin{bmatrix} 0 & 1 & 0 & 0 \\ -1 & 0 & 0 & 0 \\ 0 & 0 & 0 & 1 \\ 0 & 0 & -1 & 0 \end{bmatrix}, \mathbf{M}_2 = \begin{bmatrix} 0 & 0 & 1 & 0 \\ 0 & 0 & 0 & -1 \\ -1 & 0 & 0 & 0 \\ 0 & 1 & 0 & 0 \end{bmatrix}, \mathbf{M}_3 = \begin{bmatrix} 0 & 0 & 0 & 1 \\ 0 & 0 & 1 & 0 \\ 0 & -1 & 0 & 0 \\ -1 & 0 & 0 & 0 \end{bmatrix}$$

$$\mathbf{N}_1 = \begin{bmatrix} 0 & 1 & 0 & 0 \\ -1 & 0 & 0 & 0 \\ 0 & 0 & 0 & -1 \\ 0 & 0 & 1 & 0 \end{bmatrix}, \mathbf{N}_2 = \begin{bmatrix} 0 & 0 & 1 & 0 \\ 0 & 0 & 0 & 1 \\ -1 & 0 & 0 & 0 \\ 0 & -1 & 0 & 0 \end{bmatrix}, \mathbf{N}_3 = \begin{bmatrix} 0 & 0 & 0 & 1 \\ 0 & 0 & -1 & 0 \\ 0 & 1 & 0 & 0 \\ -1 & 0 & 0 & 0 \end{bmatrix}$$

(6)

*Example 1*: An AOD of order 8 and "*type*" (2, 2, 2, 2; 2, 2, 2, 2)

Consider an AF $\left\{ \begin{bmatrix} 1 & -1 \\ -1 & -1 \end{bmatrix}, \begin{bmatrix} 1 & 1 \\ 1 & -1 \end{bmatrix}; \begin{bmatrix} 1 & -1 \\ 1 & 1 \end{bmatrix}, \begin{bmatrix} -1 & -1 \\ 1 & -1 \end{bmatrix} \right\}$ of order 2 and "*type*" (2, 2; 2, 2). An AOD of order 8 and "*type*" (2, 2, 2, 2; 2, 2, 2, 2), can be constructed by using *Construction 1* and $\{\mathbf{M}, \mathbf{N}\}$ in (6):

$$\mathcal{A} = \begin{bmatrix} a_4 & a_4 & a_1 & -a_1 & a_2 & -a_2 & a_3 & -a_3 \\ a_4 & -a_4 & a_1 & a_1 & a_2 & a_2 & a_3 & a_3 \\ -a_1 & a_1 & a_4 & a_4 & a_3 & -a_3 & -a_2 & a_2 \\ -a_1 & -a_1 & a_4 & -a_4 & a_3 & a_3 & -a_2 & -a_2 \\ -a_2 & a_2 & -a_3 & a_3 & a_4 & a_4 & a_1 & -a_1 \\ -a_2 & -a_2 & -a_3 & -a_3 & a_4 & -a_4 & a_1 & a_1 \\ -a_3 & a_3 & a_2 & -a_2 & -a_1 & a_1 & a_4 & a_4 \\ -a_3 & -a_3 & a_2 & a_2 & -a_1 & -a_1 & a_4 & -a_4 \end{bmatrix}$$

(7)

$$\mathcal{B} = \begin{bmatrix} -b_4 & -b_4 & b_1 & -b_1 & b_2 & -b_2 & b_3 & -b_3 \\ b_4 & -b_4 & -b_1 & -b_1 & -b_2 & -b_2 & -b_3 & -b_3 \\ -b_1 & b_1 & -b_4 & -b_4 & -b_3 & b_3 & b_2 & -b_2 \\ b_1 & b_1 & b_4 & -b_4 & b_3 & b_3 & -b_2 & -b_2 \\ -b_2 & b_2 & b_3 & -b_3 & -b_4 & -b_4 & -b_1 & b_1 \\ b_2 & b_2 & -b_3 & -b_3 & b_4 & -b_4 & b_1 & b_1 \\ -b_3 & b_3 & -b_2 & b_2 & b_1 & -b_1 & -b_4 & -b_4 \\ b_3 & b_3 & b_2 & b_2 & -b_1 & -b_1 & b_4 & -b_4 \end{bmatrix}$$

By letting $x_i = a_i + jb_i$, where $j^2 = -1$, and rearranging the symbols $x_1$ to $x_4$, we obtain the O-STBC:

$$\mathbf{G8} = \begin{bmatrix} x_1^* & x_1^* & x_2 & -x_2 & x_3 & -x_3 & x_4 & -x_4 \\ x_1 & -x_1 & x_2^* & x_2^* & x_3^* & x_3^* & x_4^* & x_4^* \\ -x_2 & x_2 & x_1^* & x_1^* & x_4^* & -x_4^* & -x_3^* & x_3^* \\ -x_2^* & -x_2^* & x_1 & -x_1 & x_4 & x_4 & -x_3 & -x_3 \\ -x_3 & x_3 & -x_4^* & x_4^* & x_1^* & x_1^* & x_2^* & -x_2^* \\ -x_3^* & -x_3^* & -x_4 & -x_4 & x_1 & -x_1 & x_2 & x_2 \\ -x_4 & x_4 & x_3^* & -x_3^* & -x_2^* & x_2^* & x_1^* & x_1^* \\ -x_4^* & -x_4^* & x_3 & x_3 & -x_2 & -x_2 & x_1 & -x_1 \end{bmatrix}$$

(8)

**G8** has the form:

$$\mathcal{Q}_1 = \begin{bmatrix} \mathbf{P} & \mathbf{Q} & \mathbf{R} & \mathbf{S} \\ -\mathbf{Q} & \mathbf{P} & \mathbf{S}^* & -\mathbf{R}^* \\ -\mathbf{R} & -\mathbf{S}^* & \mathbf{P} & \mathbf{Q}^* \\ -\mathbf{S} & \mathbf{R}^* & -\mathbf{Q}^* & \mathbf{P} \end{bmatrix}$$

(9)

where $\mathbf{P} = \begin{bmatrix} x_1^* & x_1^* \\ x_1 & -x_1 \end{bmatrix}$, $\mathbf{Q} = \begin{bmatrix} x_2 & -x_2 \\ x_2^* & x_2^* \end{bmatrix}$, $\mathbf{R} = \begin{bmatrix} x_3 & -x_3 \\ x_3^* & x_3^* \end{bmatrix}$,

$\mathbf{S} = \begin{bmatrix} x_4 & -x_4 \\ x_4^* & x_4^* \end{bmatrix}$. This design is exactly the same as the one proposed in Theorem 1 in [9], hence Theorem 1 in [9] can be treated as a special case of *Construction 1*. This example also demonstrates that an AOD can be constructed from an AF.

By changing the lower-order AOD $\{\mathbf{A}, \mathbf{B}\}$ matrices or the $\{\mathbf{M}, \mathbf{N}\}$ matrices in *Construction 1*, we can obtain new and existing O-STBCs. One such example is shown in Appendix, which shows that by using another set of $\{\mathbf{A}, \mathbf{B}\}$ and $\{\mathbf{M}, \mathbf{N}\}$ matrices, Theorem 2 in [9] can be obtained. This demonstrates the generality and versatility of the proposed construction method.

*Example 2*: An AOD of order 8 and "*type*" (2, 2, 2, 2; 2, 2, 2, 2) with complex entries

Consider an AF $\left\{ \begin{bmatrix} 1 & -1 \\ -j & -j \end{bmatrix}, \begin{bmatrix} 1 & 1 \\ j & -j \end{bmatrix}; \begin{bmatrix} 1 & -1 \\ j & j \end{bmatrix}, \begin{bmatrix} -1 & -1 \\ j & -j \end{bmatrix} \right\}$ of order 2 and "*type*" (2, 2; 2, 2). The following O-STBC of order 8 and "*type*" (2, 2, 2, 2; 2, 2, 2, 2) can be constructed using *Construction 1* and $\{\mathbf{M}, \mathbf{N}\}$ in (6) (here we allow AF and AOD with complex entries, as it has been shown in [13] that AF and AOD with complex entries have the same total number of variables):

$$\mathbf{H8} = \begin{bmatrix} x_1^* & x_1^* & x_2 & -x_2 & x_3 & -x_3 & x_4 & -x_4 \\ jx_1 & -jx_1 & jx_2^* & jx_2^* & jx_3^* & jx_3^* & jx_4^* & jx_4^* \\ -x_2 & x_2 & x_1^* & x_1^* & x_4^* & -x_4^* & -x_3^* & x_3^* \\ -jx_2^* & -jx_2^* & jx_1 & -jx_1 & jx_4 & jx_4 & -jx_3 & -jx_3 \\ -x_3 & x_3 & -x_4^* & x_4^* & x_1^* & x_1^* & x_2^* & -x_2^* \\ -jx_3^* & -jx_3^* & -jx_4 & -jx_4 & jx_1 & -jx_1 & jx_2 & jx_2 \\ -x_4 & x_4 & x_3^* & -x_3^* & -x_2^* & x_2^* & x_1^* & x_1^* \\ -jx_4^* & -jx_4^* & jx_3 & jx_3 & -jx_2 & -jx_2 & jx_1 & -jx_1 \end{bmatrix}$$

(10)

Note that **H8** in (10) is the first O-STBC of order 8 that contains no zero entries first reported by us in [13]. The sub-matrix of this code does not follow condition 2 of Theorem 1 in [9], hence this code cannot be constructed directly using the proposed method in [9].

*Construction 2*: If { $\mathbf{A}_i$, $1 \leq i \leq s$ ; $\mathbf{B}_q$, $1 \leq q \leq t$ } is an AOD / AF of order $n$ with $s + t$ variables of "*type*" ($f_1, \ldots, f_s$ ; $g_1, \ldots, g_t$), and $\mathbf{N}_1 = \begin{bmatrix} 0 & 1 \\ -1 & 0 \end{bmatrix}, \mathbf{N}_2 = \begin{bmatrix} 0 & 1 \\ 1 & 0 \end{bmatrix}, \mathbf{N}_3 = \begin{bmatrix} 1 & 0 \\ 0 & -1 \end{bmatrix}$, then

$\{ \mathbf{B}_1 \otimes \mathbf{N}_1, \mathbf{A}_i \otimes \mathbf{I}_2, 1 \leq i \leq s ;$
$\mathbf{B}_1 \otimes \mathbf{N}_2, \mathbf{B}_1 \otimes \mathbf{N}_3, \mathbf{B}_q \otimes \mathbf{I}_2, 2 \leq q \leq t \}$

is an AOD / AF of order $2n$ with $s + t + 2$ variables and "*type*" ($g_1, f_1, f_2, \ldots, f_s$ ; $g_1, g_1, g_2, \ldots, g_t$).

Proof of *Construction 2*: Similar to the proof of *Construction 1*. ∎

*Proposition 2*: An AOD of order $2n$ constructed by *Construction 2* can achieve the maximum number of variables if an AOD of order $n$ with the maximum number of variables is used.

Proof of *Proposition 2*: Similar to the proof of *Proposition 1*. ∎

*Example 3*: An AOD of order 4 and "*type*" (2, 2, 2; 2, 2, 2)

Consider an AOD $\left\{ \begin{bmatrix} -1 & 1 \\ 1 & 1 \end{bmatrix}, \begin{bmatrix} 1 & 1 \\ 1 & -1 \end{bmatrix}; \begin{bmatrix} 1 & -1 \\ 1 & 1 \end{bmatrix}, \begin{bmatrix} 1 & 1 \\ -1 & 1 \end{bmatrix} \right\}$ of order 2 and "*type*" (2, 2; 2, 2), a new O-STBC for four transmit antennas, based on an AOD of order 4 and "*type*" (2, 2, 2; 2, 2, 2), denoted herein as **G4**, can be constructed:

$$\mathbf{G4} = \begin{bmatrix} x_1 - x_2^* & x_1 + x_2^* & x_3 & -x_3 \\ x_1^* + x_2 & -x_1^* + x_2 & x_3 & x_3 \\ -x_3^* & x_3^* & x_1 - x_2 & x_1 + x_2 \\ -x_3^* & -x_3^* & x_1^* + x_2^* & -x_1^* + x_2^* \end{bmatrix} \quad (11)$$

Due to the properties stated in *Proposition 1* and *Proposition 2*, all the square O-STBCs constructed so far can achieve the maximum achievable code rate of ¾. As the proposed construction methods are general, they can also be used to construct many existing O-STBCs.

## IV. CONSTRUCTION OF O-STBC FROM AOD

To quantify the benefit of having less or no zero entries in an O-STBC, we now introduce the power distribution properties of O-STBC's, and show that codes designed using Proposition 3, to be presented later in Section IV C, will have more favorable power distribution properties.

A transmitted signal with good power-distribution characteristics [6] should have:
- Low, ideally 1, peak-to-average power ratio (peak/ave)
- Low, ideally 1, average-to-minimum power ratio (ave/min)
- Low, ideally 0, probability $P_o$ that an antenna transmits "zero" (i.e. is turned off)

In general, space-time codes that have the above "ideal" attributes are "power-balanced". We next show the new O-STBCs in *Example 1* and *Example 3* have desirable attributes for practical implementation.

### A. O-STBC for Eight Transmit Antennas

Consider O-STBC designs for eight transmit antennas. Two rate-½ square O-STBCs for eight transmit antennas have been proposed in the literature: one by Tirkkonen and Hottinen [3], herein denoted as "**TH**"; the other by Tran, Seberry *et al.* [8], herein denoted as "**TS**". The **TH** code is one of the first O-STBC's for eight transmit antennas, and it is constructed from the Clifford algebraic technique. The **TS** code was designed with an aim to reduce the unused time slots (i.e. number of zeros inside the codeword). Both designs have zero coefficients in the codewords.

$$\mathbf{TH} = \begin{bmatrix} x_1 & x_2 & x_3 & 0 & x_4 & 0 & 0 & 0 \\ -x_2^* & x_1^* & 0 & -x_3 & 0 & -x_4 & 0 & 0 \\ -x_3^* & 0 & x_1^* & x_2 & 0 & 0 & -x_4 & 0 \\ 0 & x_3^* & -x_2^* & x_1 & 0 & 0 & 0 & x_4 \\ -x_4^* & 0 & 0 & 0 & x_1^* & x_2 & x_3 & 0 \\ 0 & x_4^* & 0 & 0 & -x_2^* & x_1 & 0 & -x_3 \\ 0 & 0 & x_4^* & 0 & -x_3^* & 0 & x_1 & x_2 \\ 0 & 0 & 0 & -x_4^* & 0 & x_3^* & -x_2^* & x_1^* \end{bmatrix} \quad (12)$$

$$\mathbf{TS} = \begin{bmatrix} x_1 & 0 & x_3^R + jx_2^I & x_2^R + jx_3^I & \cdots \\ 0 & x_1 & -x_2^R + jx_3^I & x_3^R - jx_2^I & \\ -x_3^R + jx_2^I & x_2^R + jx_3^I & x_1^* & 0 & \\ -x_2^R + jx_3^I & -x_3^R - jx_2^I & 0 & x_1^* & \\ -x_4^*/2 & -x_4^*/2 & -x_4^*/2 & -x_4^*/2 & \\ -x_4^*/2 & x_4^*/2 & -x_4^*/2 & x_4^*/2 & \\ -x_4^*/2 & -x_4^*/2 & x_4^*/2 & x_4^*/2 & \\ -x_4^*/2 & x_4^*/2 & x_4^*/2 & -x_4^*/2 & \\ & & & & \\ x_4/2 & x_4/2 & x_4/2 & x_4/2 \\ x_4/2 & -x_4/2 & x_4/2 & -x_4/2 \\ x_4/2 & x_4/2 & -x_4/2 & -x_4/2 \\ x_4/2 & -x_4/2 & -x_4/2 & x_4/2 \\ x_1^R - jx_3^I & x_2^* & x_3^R - jx_1^I & 0 \\ -x_2 & x_1^R + jx_3^I & 0 & x_3^R - jx_1^I \\ -x_3^R - jx_1^I & 0 & x_1^R + jx_3^I & -x_2^* \\ 0 & -x_3^R - jx_1^I & x_2 & x_1^R - jx_3^I \end{bmatrix} \quad (13)$$

In the following, the TH and TS codes will be compared to a new O-STBC, **G8**, constructed from *Example 1* in (7).

Note that half of the codeword entries in **TH** are zero, hence four of the eight transmit antennas will have to be turned off at any one time for this code, and it is constructed from an AOD of "*type*" (1, 1, 1, 1; 1, 1, 1, 1). Similarly, the code **TS** which is constructed from an AOD of "*type*" (1, 1, 1, 4; 1, 1, 1, 4) also requires one of the transmit antennas to be turned off at any one time. In contrast, the **G8** code, which is constructed from an AOD of "*type*" (2, 2, 2, 2; 2, 2, 2, 2), has no zero coefficients in the codeword and hence does not require any transmit antenna to be turned off at any one time. The fact that the **G8** code comes from an AOD that is constructed from an AF (refer to *Example 1*), shows that our construction method of an AOD from an AF plays an important role in the generation of O-STBC's without zeros in its code matrix.

The power distribution characteristics of our new ½-rate O-

STBC for eight transmit antennas are compared against existing O-STBCs, **TH** and **TS**. From Table 1, we can see that the new **G8** code has much better power-distribution characteristics than the **TH** and **TS** codes. In fact, **G8** is optimally power-balanced as it has the ideal power-distribution attributes, i.e. peak / ave = 1, ave / min = 1, and $P_o$ = 0.

Table 1 Power distribution characteristics of eight-antenna O-STBC with QPSK modulation

|  | $\frac{\text{Peak}}{\text{Ave}}$ | $\frac{\text{Ave}}{\text{Min}}$ | $P_o$ | Σ "*type*" | Σ "*type*" ≥ 16 ? |
|---|---|---|---|---|---|
| **TH** [2] | 2 | ∞ | 50% | 8 | No |
| **TS** [8] | 2 | ∞ | 12.5% | 10 | No |
| **G8** | 1 | 1 | 0 | 16 | Yes |

*B. O-STBC for Four Transmit Antennas*

Here we denote the rate-¾ O-STBCs proposed in [1] by Tarokh, Jafarkhani and Calderbank as "**TJC**", and the rate-¾ O-STBCs proposed in [2] by Ganesan and Stoica as "**GS**":

$$\mathbf{TJC} = \begin{bmatrix} x_1 & x_2 & x_3/\sqrt{2} & x_3/\sqrt{2} \\ -x_2^* & x_1^* & x_3/\sqrt{2} & -x_3/\sqrt{2} \\ \frac{x_3^*}{\sqrt{2}} & \frac{x_3^*}{\sqrt{2}} & \frac{(-x_1-x_1^*+x_2-x_2^*)}{2} & \frac{(-x_2-x_2^*+x_1-x_1^*)}{2} \\ \frac{x_3^*}{\sqrt{2}} & -\frac{x_3^*}{\sqrt{2}} & \frac{(x_2+x_2^*+x_1-x_1^*)}{2} & -\frac{(x_1+x_1^*+x_2-x_2^*)}{2} \end{bmatrix}$$ (14)

$$\mathbf{GS} = \begin{bmatrix} x_1 & 0 & x_2 & -x_3 \\ 0 & x_1 & x_3^* & x_2^* \\ -x_2^* & -x_3 & x_1^* & 0 \\ x_3^* & -x_2 & 0 & x_1^* \end{bmatrix}$$ (15)

It can be seen that the **TJC** code contains the irrational number $1/\sqrt{2}$ in some of the codeword entries. This is because the **TJC** code was formed from an AOD of "*type*" (1, 1, 2; 1, 1, 2), which means that the symbols with type "2" have twice the power as the symbols with type "1". In order to normalize the power per symbol to be the same for all symbols, the scaling factor $1/\sqrt{2}$ is needed in the **TJC**. Such multiplication operation is inconvenient to implement as compared to the addition/subtraction operation associated with just having ±1 in the code coefficients of **G4**.

On the other hand, for the **GS** code in (15) which is constructed from an AOD of "*type*" (1, 1, 1; 1, 1, 1), each transmit antennas has to be turned off once in every four code symbol durations. This shortcoming does not exist for the **G4** code if its $x_1$ and $x_2$ symbols are taken from different rotated constellations, e.g. $x_1$ from QPSK and $x_2$ from rotated-QPSK (note: constellation rotation does not affect the orthogonality of **G4**).

The power distribution characteristics of our new ¾-rate O-STBC for four transmit antennas is compared against existing O-STBCs in Table 2. From Table 2, our newly constructed **G4** code achieves better peak/ave ratio and ave/min ratio than the **GS** codes in [2], as well as the "power-balanced" version of **GS** codes in [6]. Although the **TJC** code has better power-distribution characteristics than **G4**, it contains as mentioned earlier, irrational-number coefficients inside its codeword. Due to the above reasons, **G4** code is more advantageous than the **TJC** and **GS** codes in terms of practical implementation.

Table 2 Power distribution characteristics of four-antenna O-STBC with QPSK modulation

|  | $\frac{\text{Peak}}{\text{Ave}}$ | $\frac{\text{Ave}}{\text{Min}}$ | $P_o$ | Σ "*type*" | Σ "*type*" ≥ 8 ? |
|---|---|---|---|---|---|
| **TJC** [1] | 1.33 | 1.5 | 0 | 8 | Yes |
| **GS** [2] | 1.33 | ∞ | 25% | 6 | No |
| Power-balanced **GS** #1 [6] | 3 | 3 | 0 | NA | NA |
| Power-balanced **GS** #2 [6] | 2.6 | 17.5 | 0 | NA | NA |
| **G4** | 2.28 | 2.56 | 0 | 12 | Yes |

NA: Not applicable.

*C. Guidelines for Designing Good Practical O-STBC*

From the above examples and other earlier observations, we can draw the guidelines in *Proposition 3* below, for the design of a practical O-STBC:

*Proposition 3*: To design an O-STBC without irrational number coefficients and zero entries, the following guidelines can be applied:
- To avoid irrational number coefficients inside an O-STBC codeword, an AOD having a constant "*type*" value for all variables, i.e. $f_i = g_i$ = constant $\forall i$, is desired.
- To avoid zero entries inside an O-STBC codeword, an AOD with the sum of "*type*" for all its variables equal to or greater than $2n$, i.e. $\Sigma(f_i + g_i) \geq 2n$, is desired.

Proof of *Proposition 3*: From the above examples and from 2(i) and 4(i), we know that the "*type*" of a variable in an AOD is related to the power of the corresponding transmitted symbols. If all the variables have equal "*type*", all the transmitted symbols will have equal transmission power, this will eliminate the need for power normalization for individual symbols and hence irrational number coefficients in the O-STBC codeword. Next, the "*type*" of a variable in an AOD is also related to the number of occurrence of that variable in a row of the AOD matrix. To eliminate zeros inside the codeword of an O-STBC, it is desired to have all the positions inside the codeword matrix at least filled by a symbol. So the sum of occurrence of every complex symbols must be equal to or greater than $n$, or the sum of "*type*" of all variables must be equal to or greater than $2n$ (since a complex symbol consists of two real symbols, and is represented by two variables in an AOD). However, this is only a necessary, but not sufficient, condition to eliminate the zeros inside an O-STBC codeword.
∎

Both Table 1 and Table 2 also show that codes that satisfy the second design guideline of *Proposition 3*, i.e. sum of "*type*" for all variables greater than or equal to $2n$, will achieve $P_o = 0$. It can be easily shown that all the new O-STBCs proposed in this paper have exactly the same code rate, coding gain and diversity gains as the existing O-STBCs [1,2,6], hence their practical advantages (better power distribution properties) do not come with rate or performance penalty.

## V. CONCLUSION

In this paper, we have proposed two new systematic ways to construct higher-order AOD's or AF's from lower-order AOD's or AF's. We found that the "*type*" parameter in an AOD plays an important role in shaping the power distribution and code coefficient characteristics of the O-STBC. From these insights, we propose two guidelines on how to select the "*type*" parameter of AOD for constructing new O-STBC with favorable implementation attributes. New square O-STBCs for four and eight transmit antennas are constructed using the proposed AOD construction methods and design guidelines. They are shown to possess better power distribution or rational code coefficient characteristics than the existing O-STBC. In particular, one of the newly constructed O-STBCs for eight transmit antennas is found to be the first to achieve the optimal power distribution properties. Interestingly, new AOD's that lead to good practical O-STBC's can be constructed from lower-ordered AF's. The proposed construction method is general and inclusive of some of the construction methods proposed in the literature.

### APPENDIX: EQUIVALENCE OF THEOREM 1 AND 2 OF [9]

We will show that Theorem 1 and 2 of [9] can be unified by our proposed construction method.

*Example 4*: An AOD of order 8 and "*type*" (2, 2, 2, 2; 2, 2, 2, 2) with new set of {**M**, **N**}
Use the set of {**M**, **N**} as shown below:

$$\mathbf{M}_1 = \begin{bmatrix} 0 & 1 & 0 & 0 \\ -1 & 0 & 0 & 0 \\ 0 & 0 & 0 & -1 \\ 0 & 0 & 1 & 0 \end{bmatrix}, \mathbf{M}_2 = \begin{bmatrix} 0 & 0 & 1 & 0 \\ 0 & 0 & 0 & 1 \\ -1 & 0 & 0 & 0 \\ 0 & -1 & 0 & 0 \end{bmatrix}, \mathbf{M}_3 = \begin{bmatrix} 0 & 0 & 0 & 1 \\ 0 & 0 & -1 & 0 \\ 0 & 1 & 0 & 0 \\ -1 & 0 & 0 & 0 \end{bmatrix}$$

$$\mathbf{N}_1 = \begin{bmatrix} 0 & 1 & 0 & 0 \\ -1 & 0 & 0 & 0 \\ 0 & 0 & 0 & 1 \\ 0 & 0 & -1 & 0 \end{bmatrix}, \mathbf{N}_2 = \begin{bmatrix} 0 & 0 & 1 & 0 \\ 0 & 0 & 0 & -1 \\ -1 & 0 & 0 & 0 \\ 0 & 1 & 0 & 0 \end{bmatrix}, \mathbf{N}_3 = \begin{bmatrix} 0 & 0 & 0 & 1 \\ 0 & 0 & 1 & 0 \\ 0 & -1 & 0 & 0 \\ -1 & 0 & 0 & 0 \end{bmatrix}$$
(16)

And the AF $\left\{ \begin{bmatrix} 1 & -1 \\ -1 & -1 \end{bmatrix}, \begin{bmatrix} 1 & 1 \\ 1 & -1 \end{bmatrix}; \begin{bmatrix} 1 & -1 \\ 1 & 1 \end{bmatrix}, \begin{bmatrix} -1 & -1 \\ 1 & -1 \end{bmatrix} \right\}$ of order 2 and "*type*" (2, 2; 2, 2) on *Construction 1*, we obtain the followings O-STBC F8, which has the form in $\mathcal{Q}_2$:

$$\mathbf{F8} = \begin{bmatrix} x_1^* & x_1^* & x_2 & -x_2 & x_3 & -x_3 & x_4 & -x_4 \\ x_1 & -x_1 & x_2^* & x_2^* & x_3^* & x_3^* & x_4^* & x_4^* \\ -x_2 & x_2 & x_1^* & x_1^* & -x_4^* & x_4^* & x_3^* & -x_3^* \\ -x_2^* & -x_2^* & x_1 & -x_1 & -x_4 & -x_4 & x_3 & x_3 \\ -x_3 & x_3 & x_4^* & -x_4^* & x_1^* & x_1^* & -x_2^* & x_2^* \\ -x_3^* & -x_3^* & x_4 & x_4 & x_1 & -x_1 & -x_2 & -x_2 \\ -x_4 & x_4 & -x_3^* & x_3^* & x_2^* & -x_2^* & x_1^* & x_1^* \\ -x_4^* & -x_4^* & -x_3 & -x_3 & x_2 & x_2 & x_1 & -x_1 \end{bmatrix}$$

$$\Rightarrow \mathcal{Q}_2 = \begin{bmatrix} \mathbf{P} & \mathbf{Q} & \mathbf{R} & \mathbf{S} \\ -\mathbf{Q} & \mathbf{P} & -\mathbf{S}^* & \mathbf{R}^* \\ -\mathbf{R} & \mathbf{S}^* & \mathbf{P} & -\mathbf{Q}^* \\ -\mathbf{S} & -\mathbf{R}^* & \mathbf{Q}^* & \mathbf{P} \end{bmatrix}$$
(17)

Although it may not be obvious, it can be shown that by interchanging the first two and the last two columns and negating the last two rows of $\mathcal{Q}_2$, it becomes exactly the Theorem 2 proposed in [9].

$$\begin{bmatrix} 1 & 0 & 0 & 0 \\ 0 & 1 & 0 & 0 \\ 0 & 0 & -1 & 0 \\ 0 & 0 & 0 & -1 \end{bmatrix} \mathcal{Q}_2 \begin{bmatrix} 0 & 0 & 1 & 0 \\ 0 & 0 & 0 & 1 \\ 1 & 0 & 0 & 0 \\ 0 & 1 & 0 & 0 \end{bmatrix} = \begin{bmatrix} \mathbf{R} & \mathbf{S} & \mathbf{P} & \mathbf{Q} \\ -\mathbf{S}^* & \mathbf{R}^* & -\mathbf{Q} & \mathbf{P} \\ -\mathbf{P} & \mathbf{Q}^* & \mathbf{R} & -\mathbf{S}^* \\ -\mathbf{Q}^* & -\mathbf{P} & \mathbf{S} & \mathbf{R}^* \end{bmatrix}$$

= Thoeream 2 of [9]

We have also shown earlier in *Example 1* that $\mathcal{Q}_1$ and Theorem 1 of [9] are of the same design. Hence the proposed *Construction 1* unifies both Theorem 1 and 2 of [9] by using different set of {**M**, **N**}. ■